\documentclass[a4paper,11pt]{article}
\usepackage{aaskaiid}
\usepackage{natbib}
\usepackage{orcidlink}

\title{The Emerging Population of High-energy Emitting Radio Galaxies}
\ShortTitle{High-energy radio galaxies}

\author[1]{G. Bruni\orcidlink{0000-0002-5182-6289}}
\ShortName{Bruni et al.}
\author[2]{V. S. Paliya}
\author[2,3]{D. J. Saikia}
\author[4]{L. Bassani}
\author[5]{R. D. Baldi\orcidlink{0000-0002-1824-0411}}
\author[4]{E. Bronzini\orcidlink{0000-0001-8378-4303}}
\author[5]{F. D'Ammando}
\author[6]{S. del Palacio}
\author[7]{M. Kadler}
\author[8]{Y.~Y.~Kovalev\orcidlink{0000-0001-9303-3263}}
\author[9]{D. V. Lal}
\author[5]{G. Migliori}
\author[10]{B. Mingo\orcidlink{0000-0001-5649-938X}}
\author[11]{J. Moldon\orcidlink{0000-0002-8079-7608}}
\author[12,13]{L. Ostorero\orcidlink{0000-0003-3983-5980}}
\author[1]{F. Panessa}
\author[14,15]{M. Persic}
\author[5]{I. Prandoni\orcidlink{0000-0001-9680-7092}}
\author[7,8]{L. Ricci}
\author[16,17,8]{T. Savolainen}
\author[18]{F. Shankar}
\author[19]{F. Tavecchio}
\author[20]{E. Traianou}
\author[5,21]{F. Ubertosi\orcidlink{0000-0001-5338-4472}}
\author[5]{T. Venturi}

\affiliation[1]{INAF - Istituto di Astrofisica e Planetologia Spaziali, via del Fosso del Cavaliere 100, I-00133 Rome, Italy}
\affiliation[2]{Inter-University Centre for Astronomy and Astrophysics (IUCAA), SPPU Campus, Pune 411 007, India}
\affiliation[3]{Fakult\"at f\"ur Physik, Universit\"at Bielefeld, 33501 Bielefeld, Germany}
\affiliation[4]{INAF - Osservatorio di Astroﬁsica e Scienza dello Spazio, via Piero Gobetti 93/3, I-40129 Bologna, Italy}
\affiliation[5]{INAF - Istituto di Radioastronomia, Via Gobetti 101, I-40129 Bologna, Italy}
\affiliation[6]{Department of Space, Earth and Environment, Chalmers University of Technology, SE-412 96 Gothenburg, Sweden}
\affiliation[7]{Julius-Maximilians-Universit{\"a}t W{\"u}rzburg, Fakult{\"a}t für Physik und Astronomie, Institut für Theoretische Physik und Astrophysik, Lehrstuhl für Astronomie, Emil-Fischer-Str. 31, D-97074 W{\"u}rzburg, Germany}
\affiliation[8]{Max Planck Institute for Radio Astronomy, Auf dem Huegel 69, 53121 Bonn, Germany}
\affiliation[9]{National Centre for Radio Astrophysics - Tata Institute of Fundamental Research, , Post Box 3, Ganeshkhind P.O., Pune 411007, India}
\affiliation[10]{Centre for Astrophysics Research, Department of Physics, Astronomy and Mathematics, University of Hertfordshire, College Lane, Hatfield AL10 9AB, UK}
\affiliation[11]{Instituto de Astrofísica de Andalucía, Consejo Superior de Investigaciones Científicas (CSIC), Glorieta de la Astronomía s/n, E-18008 Granada, Spain}
\affiliation[12]{Dipartimento di Fisica, Università degli Studi di Torino, via P. Giuria 1, I-10125 Torino, Italy}
\affiliation[13]{INFN - Istituto Nazionale di Fisica Nucleare, Sezione di Torino, Via P. Giuria 1, I-10125 Torino, Italy}
\affiliation[14]{INAF - Osservatorio Astronomico di Padova, vicolo dell’Osservatorio 5, I-35122 Padova, Italy}
\affiliation[15]{INFN - Sezione di Trieste (Gruppo collegato di Udine), I-34127 Trieste, Italy}
\affiliation[16]{Aalto University Department of Electronics and Nanoengineering, PL\,15500, FI-00076 Aalto, Finland}
\affiliation[17]{Aalto University Metsähovi Radio Observatory, Metsähovintie 114, FI-02540 Kylmälä, Finland}
\affiliation[18]{School of Physics and Astronomy, University of Southampton, Highfield, Southampton, SO17 1BJ, UK}
\affiliation[19]{INAF - Osservatorio Astronomico di Brera, Via E. Bianchi 46, I-23807 Merate, Italy}
\affiliation[20]{Interdisciplinary Center for Scientific Computing (IWR), Mathematikon Im Neuenheimer Feld 205, 69120 Heidelberg, Germany}
\affiliation[21]{Dipartimento di Fisica e Astronomia, Università di Bologna, via Gobetti 93/2, I-40129 Bologna, Italy}

\abstract{High-energy emission from radio galaxies provides a unique laboratory to study the connection between accretion, jet formation, and particle acceleration in active galactic nuclei (AGN). 
The recent detection of $\gamma$-ray emission from misaligned radio galaxies - including Compact Symmetric Objects (CSOs), FR~0, FR~I/II, and even Giant Radio Galaxies (GRGs) - has shown that efficient particle acceleration is not limited to blazars, but occurs throughout the full radio-loud AGN population. 
This finding supports a unifying framework where leptonic synchrotron, synchrotron self-Compton (SSC), and external inverse-Compton (EIC) processes coexist across multiple spatial scales, from the inner jet and corona to the extended lobes, possibly with a hadronic contribution in dense environments.

The Square Kilometre Array (SKA) will be pivotal in advancing this field. 
SKA1-Low will detect and characterize diffuse, low-surface-brightness emission tracing aged plasma and jet duty cycles. 
SKA1-Mid will enable high-resolution spectral and polarimetric studies of compact jets and nuclear regions, while SKA-VLBI will connect parsec- to kiloparsec-scale structures, identifying the exact sites of high-energy dissipation. 
In synergy with forthcoming high-energy missions such as \emph{NewAthena} and CTAO, SKA will provide the first spatially resolved, multi-scale view of particle acceleration and energy release in misaligned AGN, unveiling the physical link between the central engine and its large-scale feedback on the host galaxy evolution.}

\begin{document}
\maketitle

%\section{Possible themes}
%\begin{itemize}
%    \item The discovery of faint/distant/previously unresolved radio galaxies with SKA
%    \item SED studies through the wide SKA frequency coverage 
%    \item The detailed view of the lobes and cores with SKA-VLBI (Low and Mid): modeling the expected high energy emission
%    \item High energy emission from CSO: opportunities offered by SKA-VLBI and synergies with VHE facilities.
%\end{itemize}

\section{The high-energy counterpart of the radio sky}

The detection of radio galaxies in the $\gamma$–ray domain has profoundly changed our understanding of relativistic jets seen at large inclination angles.  
The Large Area Telescope (LAT) on board the \textit{Fermi Gamma-ray Space Telescope} has shown that the extragalactic $\gamma$-ray sky is dominated by blazars, i.e. radio–loud active galactic nuclei (AGN) whose jets are closely aligned with the observer’s line of sight.  
A small fraction -- about two per cent -- of the LAT sources were instead classified as misaligned AGN (MAGN; \citealt{Abdo2010,Ajello2020}).  In this chapter, we focus on the subset of MAGN that are radio-loud, i.e. radio galaxies, which provide unique laboratories for high-energy processes without the strong Doppler boosting that characterises blazars.

Historically, radio galaxies were divided by \citet{Fanaroff1974} into two morphological classes: edge–darkened FR\,I and edge–brightened FR\,II sources \citep{Bernie2021,LalBernie2021}.  
The former were found to be generally associated with low–excitation optical spectra and radiatively inefficient accretion, while the latter usually host powerful, radiatively efficient nuclei \citep{Best2012}. However, recent LOFAR studies have shown that this picture is far more complex: large statistical samples reveal substantial populations of low-luminosity FR\,II and high-luminosity FR\,I systems, demonstrating that morphology does not map cleanly onto radio power or accretion mode \citep{Hardcastle2019,Mingo2019,Mingo2022,hardcastle26SKA}.
FR\,I and FR\,II sources overlap widely in luminosity, and both classes are predominantly LERG-dominated, indicating that jet dynamics and environment—rather than accretion regime alone—play the primary role in shaping the observed radio structures \citep{Mingo2019,Mingo2022}.   
Moreover, high–resolution observations soon indicated that the canonical FR\,I/FR\,II dichotomy was not exhaustive.  
Large-area optical and radio surveys such as the Sloan Digital Sky Survey (SDSS), the NRAO VLA Sky Survey (NVSS; \citealt{Condon1998}) and the Faint Images of the Radio Sky at Twenty centimetres (FIRST; \citealt{Becker1995}) unveiled a dominant population of compact radio–loud AGN lacking extended ($\gtrsim$ kpc) emission (\citealt{Baldi2009,Best2012}).  
This new class, termed Fanaroff–Riley type 0 (FR\,0; \citealt{Ghisellini2011}), shares the host and nuclear properties of FR\,I galaxies, but exhibits a dramatic deficit of large–scale jets \citep{baldi23,baldi26SKA}. Finally, early \textit{Fermi} analyses revealed that most $\gamma$–ray–detected radio galaxies belonged to the FR\,I subclass (e.g. Cen~A, NGC~1275, M87, \citealt{Abdo2010}), suggesting that low–power jets can accelerate particles to high energies.

\subsection{Classical misaligned AGN: FR\,I and FR\,II sources}

%While FR\,0s populate the low–power end of the radio–loud sequence, the traditional FR\,I and FR\,II galaxies remain crucial to understanding jet physics across scales. 

Systematic studies of \textit{Fermi}–LAT data have established that both FR\,I and FR\,II subclasses can produce variable GeV emission, occasionally extending into the TeV band.  
For FR\,Is, one–zone synchrotron self-Compton (SSC) models generally reproduce the global spectral energy distribution (SED), although the observed fast variability and high–energy cutoffs often require multi–zone or spine–sheath jet structures \citep{ghisellini2005,tavecchio2008}.
The detection of the FR\,II galaxy IGR~J18249$-$3243 at GeV energies \citep{Bruni2022} demonstrated that powerful edge–brightened systems can also emit efficiently in the $\gamma$–ray domain.  
Broad–band modelling of this object revealed that inverse-Compton (IC) scattering in the radio lobes, rather than the beamed core, can dominate the observed high–energy output -- a scenario reminiscent of the lobes of Fornax~A and Cen~A (see section 1.1.1).  
Such findings emphasise the contribution of extended, non–thermal plasma regions to the total energy budget of powerful radio galaxies \citep{LalRao04,LalHardcastle2008}.

%\subsection{Giant Radio Galaxies at high energies}

On a larger scale, giant radio galaxies (GRGs), with projected sizes exceeding $\sim$0.7 Mpc, trace the long–term evolution of relativistic jets and their interaction with the surrounding medium.  
Although their large viewing angle makes Doppler boosting negligible, \citet{Paliya2025} recently reported a systematic search for $\gamma$–ray–emitting GRGs, identifying sixteen examples -- half of them newly classified -- using low-frequency surveys such as the LOFAR Two-metre Sky Survey (LoTSS; \citealt{Shimwell2017}) and the Rapid ASKAP Continuum Survey (RACS; \citealt{McConnell2020}). 
The $\gamma$–ray spectral properties of GRGs are indistinguishable from those of smaller–scale misaligned AGN, suggesting a common origin for $\gamma$–ray radiation.  
Interestingly, many of these giants show hybrid or diffuse morphologies with FR\,I–like lobes, supporting the idea that particle acceleration and inverse-Compton emission persist over Mpc scales even in aged or recurrent jets (\citealt{Bruni2020,Dabhade2020}).

%%%%%%%%%%%%%%%%%%%%%%%%%%%%%%%%%%%%%%%%%%%%%%%%%%%%%%%%%%%%%%%%%%%%

\begin{figure}
    \centering
    \includegraphics[width=0.8\linewidth]{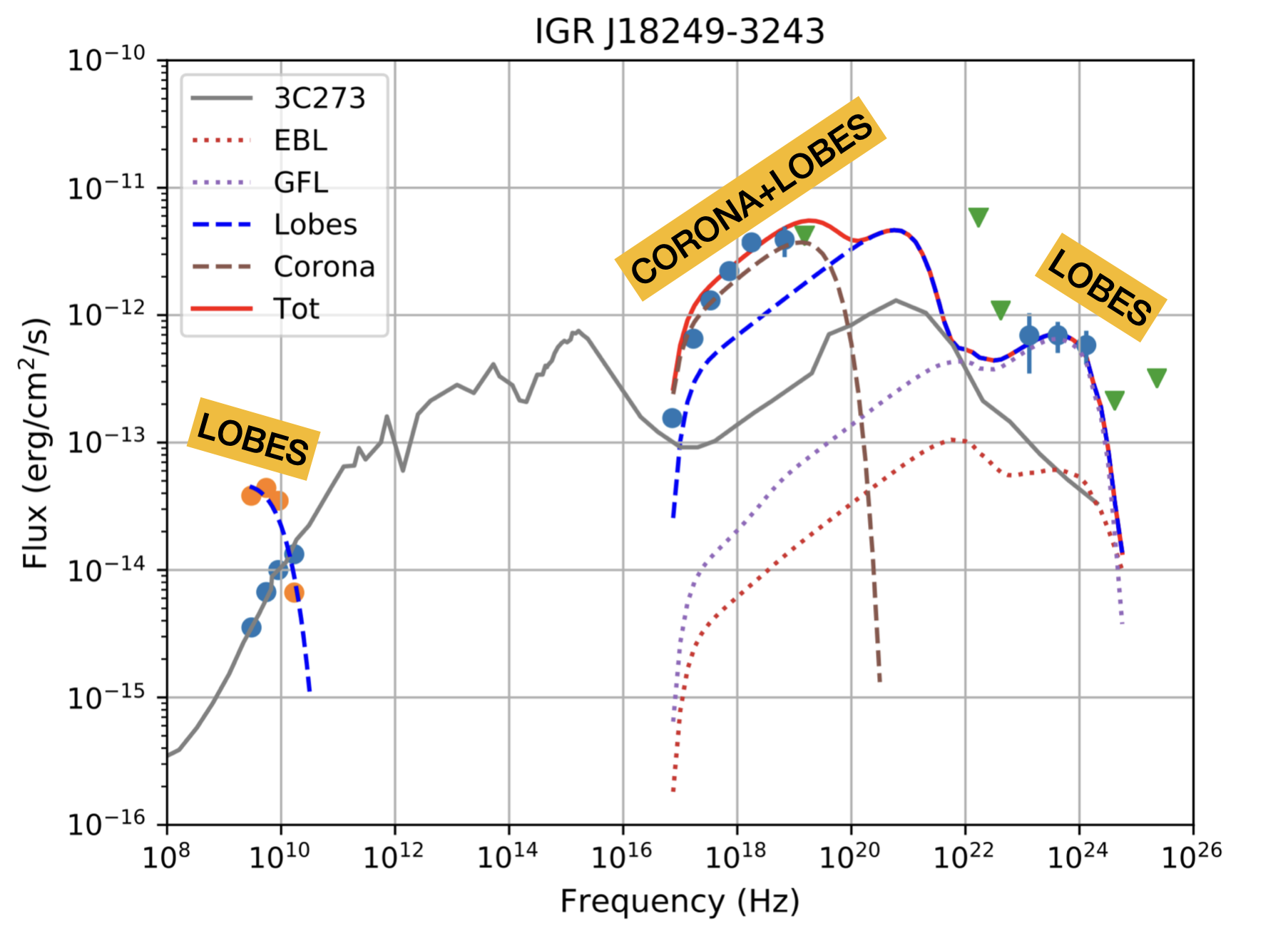}
    \caption{Broadband SED of IGR~J18249$-$3243 (adapted from \citealt{Bruni2022}).  Orange and blue symbols in the radio frequency range represent the lobe and core flux densities, respectively.  The grey line shows the average spectrum of the blazar 3C\,273 normalized to the 5\,GHz core flux of the target, serving as a reference for a typical beamed AGN spectrum. Dashed curves mark the individual model components (dotted for lobe sub-components), and the red solid line indicates the total emission.
    }
    \label{fig:IGR}
\end{figure}

%%%%%%%%%%%%%%%%%%%%%%%%%%%%%%%%%%%%%%%%%%%%%%%%%%%%%%%%%%%%%%%%%%%%

\subsection{The emergence of $\gamma$–ray–emitting FR\,0s}

The first evidence of an FR\,0 galaxy in the GeV sky was reported by \citet{Grandi2016}, who associated the source 3FGL~J1330.0$-$3818 with Tol~1326$-$379.  
This discovery demonstrated that even the most compact members of the radio–loud family can produce high–energy photons.  
Subsequent analyses with extended \textit{Fermi}-LAT data sets confirmed and expanded this result.  
Using more than a decade of LAT observations, \citet{Paliya2021} detected three additional FR\,0s above 1 GeV and obtained a statistically significant stacked signal from the whole FR0CAT sample \citep{baldi18}, proving that the population as a whole can emit in $\gamma$–rays.  
Later, \citet{Pannikkote2023} performed the complementary search, starting from the LAT AGN catalogue and cross–matching the $\gamma$-ray sources with optical and radio surveys, and identified seven bona fide $\gamma$–ray–emitting FR\,0s potentially coincident with a $\gamma$-ray counterpart.  
Their nuclear spectra are typical of low–excitation galaxies, their radio morphologies remain unresolved even at sub–arcsecond resolution \citep{baldi19a}, and the inferred Eddington ratios ($L/L_{\mathrm{Edd}} < 10^{-2}$) are consistent with radiatively inefficient accretion flows \citep{baldi19b}.
The $\gamma$–ray spectra of FR\,0s are remarkably similar to those of FR\,I galaxies, implying that comparable radiative mechanisms, likely synchrotron self–Compton (SSC) emission from mildly beamed jets, are at work \citep{torresi18,baldi19}.
Given their sheer abundance, FR\,0s have been proposed as promising contributors to the diffuse neutrino background and ultra–high–energy cosmic–ray production \citep{Tavecchio2018,Paliya2021}.

\subsection{Compact Symmetric Objects and the onset of high-energy jet activity}

A distinct and particularly intriguing subclass within the misaligned AGN family is represented by Compact Symmetric Objects (CSOs). These sources are characterized by symmetric, sub–kiloparsec radio morphologies dominated by mini-lobes on both sides of a weak or unresolved core \citep{Wilkinson1994,Readhead1996}. With linear sizes of a few parsecs to a few hundred parsecs and kinematic ages typically below a few thousand years, CSOs are considered the earliest observable phase of radio galaxy evolution. They are often associated with gigahertz-peaked spectrum (GPS) sources, whose convex radio spectra peak near 1~GHz as a consequence of synchrotron self-absorption or free–free absorption in the dense circumnuclear medium \citep{ODea1998}. Their compactness and high ambient photon density make them prime candidates for non-thermal $\gamma$-ray emission through IC scattering of nuclear and interstellar radiation fields \citep{Stawarz2008,Kino2009}.

%\citet{Migliori2014} first explored the theoretical $\gamma$-ray detectability of CSOs and other young radio sources, showing that compact lobes expanding within the host galaxy could upscatter optical–infrared photons from the accretion disk and dusty torus to GeV energies. The predicted $\gamma$-ray luminosities, ranging from $10^{41}$ to $10^{46}$ erg s$^{-1}$, depend sensitively on the jet power and the ambient photon density. However, early searches with the \textit{Fermi}-LAT did not yield significant detections, suggesting that only the youngest and nearest CSOs would be bright enough in the GeV band. This prediction was confirmed by the breakthrough detection of PKS~1718$-$649 \citep{Migliori2016}, which remains one of the most compact $\gamma$-ray–emitting AGN known, with a linear size of $\sim$2~pc and a kinematic age of $\sim$100~years. Its steep $\gamma$-ray spectrum ($\Gamma_{\gamma} \simeq 2.9$, with $F_{\gamma}(E) \propto E^{-\Gamma_{\gamma}}$) and isotropic, non-variable emission point to an origin in the compact radio lobes rather than in a beamed jet, consistent with the IC scenario originally proposed by \citet{Stawarz2008}.

The first theoretical expectations for $\gamma$-ray emission from genuinely young radio galaxies were developed by \citet{Stawarz2008}, who showed that compact, slowly expanding CSO lobes embedded in the dense circumnuclear environment can efficiently upscatter IR--UV photons from the dusty torus and the accretion disk to high energies. This framework was subsequently applied to X-ray–detected CSOs by \citet{Ostorero2010}, providing the first broadband modelling of unbeamed GPS/CSO radio galaxies in the high-energy domain. In parallel, \citet{Migliori2014} explored the $\gamma$-ray detectability of young radio sources of the GPS/CSS class, but focusing on radio \emph{quasars} whose high-energy emission is expected to arise from mildly beamed jets viewed at moderate angles (10--50 degrees), rather than from expanding lobes. 

Early \textit{Fermi}-LAT searches did not yield significant detections in either class, suggesting that only the youngest and nearest CSOs would be bright enough in the GeV band. This prediction was confirmed by the breakthrough detection of PKS~1718$-$649 \citep{Migliori2016}, one of the most compact $\gamma$-ray–emitting AGN known, with a linear size of $\sim$2\,pc and a kinematic age of $\sim$100\,yr. Its steep $\gamma$-ray spectrum ($\Gamma_{\gamma} \simeq 2.9$, with $F_{\gamma}(E)\propto E^{-\Gamma_{\gamma}}$) and isotropic, non-variable emission strongly favour an origin in the compact radio lobes, fully consistent with the inverse-Compton scenario originally proposed by \citet{Stawarz2008}.

The population of $\gamma$-ray–detected CSOs has since grown slowly but steadily. VLBI and multi-wavelength follow-ups of \textit{Fermi}-LAT sources revealed a handful of nearby examples, such as TXS~0128+554 \citep{Lister2020}, NGC~3894 \citep{Principe2020}, and most recently NGC~4278 \citep{Cao,Bronzini} and DA~362 \citep{Swain2025}. Together with PKS~1718$-$649 (NGC6328), these sources share compact, two-sided morphologies and subluminal jet advance speeds ($v_{\rm app} \lesssim 0.3c$), confirming their youth and large viewing angles. Their $\gamma$-ray luminosities ($L_{\gamma} \sim 10^{42}$–$10^{44}$~erg~s$^{-1}$) and steep photon indices ($\Gamma_{\gamma} \simeq 2.5$–3.0) are overall comparable to those of FR~I radio galaxies. Remarkably, DA~362 exhibited the first recorded $\gamma$-ray flare from a CSO, implying that transient jet–ISM interactions or intermittent accretion episodes might contribute to the high-energy variability \citep{Swain2025}. Such episodic activity is consistent with the ``restarting'' nature of several young radio galaxies and could mark recurrent phases of jet launching during the early AGN life cycle.

From a physical standpoint, the $\gamma$-ray emission in CSOs can arise either from non-relativistic lobes via IC scattering of host galaxy photon fields or, in some cases, from mildly beamed cores if the inner jets are temporarily re-energized \citep{Lister2020}. The contribution of hadronic interactions in dense nuclear environments has also been proposed \citep{KinoAsano2011}. Because of their compactness, CSOs are expected to evolve rapidly, fading below the LAT sensitivity as their lobes expand and the external photon energy density declines. Their detection thus offers a unique glimpse into the onset of jet-driven feedback and particle acceleration on sub-kiloparsec scales. As highlighted by \citet{Swain2025}, systematic searches combining VLBI-selected CSO catalogs with deep \textit{Fermi}-LAT exposures are now unveiling a small but growing population of these young, high-energy radio galaxies, bridging the gap between compact GPS sources and extended FR~I/FR~II systems.

\subsection{Where do we stand?}

After more than seventeen years of continuous \textit{Fermi} monitoring, the panorama of high–energy misaligned AGN has expanded from a handful of bright FR\,Is to an increasingly diversified population encompassing FR\,0s, FR\,IIs, and GRGs.  
Across this sequence, the $\gamma$--ray emission is typically characterised by soft photon indices ($\Gamma_\gamma \simeq 2.1$--$2.6$), although harder spectra have also been observed in some cases, and isotropic luminosities spanning $10^{42}$--$10^{45}$\,erg\,s$^{-1}$. 
These properties are consistent with leptonic radiative processes occurring in relativistic jets viewed at large inclination angles, where the Doppler boosting is modest (even though the intrinsic jet bulk speeds can reach highly relativistic values). 
The detection of compact, low–power FR\,0s bridges the gap between classical radio galaxies and the abundant, faint radio–loud nuclei in the local Universe, while the discovery of $\gamma$–ray–bright giants connects the opposite extreme of jet evolution \citep{ye26}.  
Together, these results delineate a unified picture in which $\gamma$–ray emission is a ubiquitous, though angle–dependent, feature of jetted AGN, from newborn compact sources to the most extended radio structures.  
Future facilities such as the Cherenkov Telescope Array Observatory and the SKA will be instrumental in resolving the spatial origin of high–energy radiation and constraining its role in cosmic–ray and neutrino production.

%%%%%%%%%%%%%%%%%%%%%%%%%%%%%%%%%%%%%%%%%%%%%%%%%%%%%%%%%%%%%%%%
\begin{table}
\centering
%\footnotesize
\setlength{\tabcolsep}{7.0pt}
\caption{Summary of the main classes of misaligned radio galaxies discussed in this work.}
\label{tab:misaligned_agn}
\begin{tabular}{l l l p{5.5cm}}
\hline
Class & Typical size & Typical nucleus & Dominant $\gamma$-ray mechanism \\
\hline
CSO    & $\lesssim$1 kpc   & LERG          & IC in compact lobes (SSC possible) \\
FR\,0  & $\lesssim$ 5 kpc & LERG          & SSC in compact jets \\
FR\,I  & 10-100 kpc      & LERG          & SSC (jets) $+$ lobe IC (CMB/EBL) \\
FR\,II & 0.1--1 Mpc  & LERG / HERG          & Lobe IC (CMB/EBL/GLF) \\
GRG    & $\gtrsim$ 0.7\,Mpc & LERG / HERG   & Lobe IC on Mpc scales \\
\hline
\end{tabular}

\vspace{1ex}
{\footnotesize{\textit{Notes.} HERG / LERG = high-excitation / low-excitation radio galaxy; SSC = synchrotron self-Compton;\\
IC = inverse-Compton; CMB = cosmic microwave background; EBL = extragalactic background light; \\
GLF = galactic light field.}}
\end{table}
%%%%%%%%%%%%%%%%%%%%%%%%%%%%%%%%%%%%%%%%%%%%%%%%%%%%%%%%%%%%%%%%

\section{High-Energy Emission Processes in Misaligned Radio Galaxies}

\begin{figure}
    \centering
    \includegraphics[width=1\textwidth]{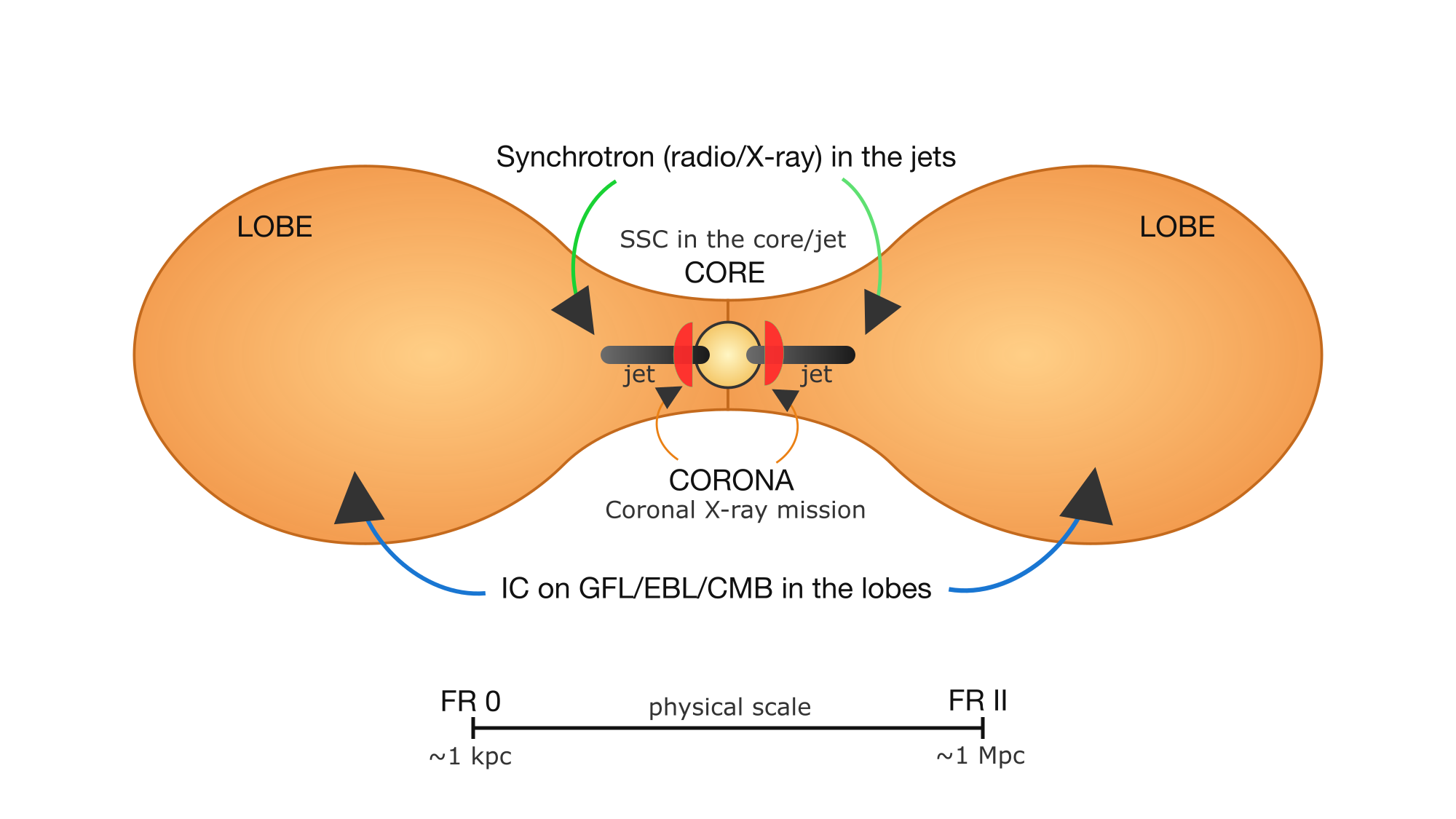}
    \caption{Sketch of the different radiative processes at play in radio galaxies: synchrotron and synchrotron self-Compton are predominant in the inner region of the jet, along with coronal emission in the X-ray band. In the lobes, inverse Compton off Galactic Foreground Light (GFL), External Background Light (EBL), and Cosmic Microwave Background (CMB) is present. The typical physical scale (bottom ruler) spans from $\sim$1 kpc for FR\,0 (or even sub-kpc for CSO) to $\sim$1 Mpc for FR\,II.}
    \label{fig:sketch}
\end{figure}

The detection of $\gamma$-rays from misaligned radio-loud AGN -- and in particular from the recently emerging populations of FR\,0s, FR\,Is, FR\,IIs radio galaxies, Giant Radio Galaxies (GRGs), and Compact Symmetric Objects (CSOs) -- provides a crucial window into the radiative mechanisms operating in relativistic jets observed at large viewing angles \citep{LalRao04,Lalremnant2021}.  
Although these sources share a common jet origin, the high-energy emission arises from a variety of physical regions and processes, whose relative contributions depend on jet power, orientation, and evolutionary stage. Below, we provide a brief overview of the radiative processes operating in high-energy–emitting radio galaxies, which are summarized in the schematic representation shown in Fig. \ref{fig:sketch}, and in Tab. \ref{tab:misaligned_agn}.

\subsection{Synchrotron and Synchrotron Self-Compton in Compact Jets}

For most of the $\gamma$-ray–detected misaligned AGN, including FR\,0s and low-power FR\,Is, the dominant mechanism is believed to be SSC scattering in the mildly relativistic jet.  
In this scenario, the same population of electrons responsible for the radio–to–X-ray synchrotron emission upscatters its own synchrotron photons to the GeV band.  
Because FR\,0s lack strong beaming and display soft photon indices ($\Gamma_{\gamma} \simeq 2.1$–2.6) and moderate luminosities ($L_{\gamma} \lesssim 10^{44-45}$ erg\,s$^{-1}$), one-zone SSC models provide a natural explanation for their observed $\gamma$-ray output \citep{Paliya2021,Pannikkote2023,2024ApJ...971...84K}.  
However, reproducing both the spectral shape and the flux level often requires more complex geometries, such as stratified (spine–sheath) flows or multi-zone structures \citep{Paliya2021,2023ApJ...955L..41B}.  
These allow for internal Compton scattering between slower and faster jet components, boosting the high-energy efficiency without invoking strong Doppler factors.

\subsection{External Inverse Compton of Host Galaxy Photon Fields}

In more powerful FR\,IIs and some GRGs, IC scattering of external photon fields, from the broad-line region, dusty torus, or host-galaxy starlight (Galactic Foreground Light, GFL), may supplement the SSC component \citep{LalKraft2013,LalKraft2010}.  
However, in FR\,0s and FR\,Is, which generally host low-excitation nuclei (LERG type) and accrete at $L/L_{\mathrm{Edd}} < 10^{-2}$, the radiation fields external to the jet are comparatively weak.  
Thus, External Inverse Compton (EIC) emission contributes only modestly, if at all, to the total $\gamma$-ray luminosity. 
Nevertheless, photons from the host galaxy can still serve as IC targets, particularly in sources that exhibit small-scale extended structures.  
This transitional regime between pure SSC and extended-lobe IC may connect FR\,0s to young CSOs and large-scale FR\,IIs.

%%%%%%%%%%%%%%%%%%%%%%%%%%%%%%%%%%%%%%%%%%%%%%%%%%%%%%%%%%%

%%%%%%%%%%%%%%%%%%%%%%%%%%%%%%%%%%%%%%%%%%%%%%%%%%%%%%%%%%%
\subsection{External Inverse Compton in the Extended Lobes}

At larger scales, non-thermal electrons in the radio lobes can upscatter cosmic microwave background (CMB) and extragalactic background light (EBL) photons to X-ray and $\gamma$-ray energies.  
This process, first established in nearby misaligned AGN such as Cen\,A, has now been firmly detected in the FR\,II galaxy IGR~J18249$-$3243 \citep{Bruni2022}.  
In this source, the high-energy spectrum is dominated by IC emission from the lobes rather than by the nuclear jet, illustrating that large-scale, isotropic components can rival the core in $\gamma$-ray power.  
For GRGs, whose lobes extend beyond $\sim$Mpc scales, this mechanism is expected to dominate the $\gamma$-ray output, consistent with the population statistics presented by \citet{Paliya2025}.
%
%\subsubsection{Leptonic modelling of lobe IC emission}
%\label{sec:lobeic}
%
A consistent framework for interpreting the high-energy emission from radio-galaxy lobes has recently emerged from a sequence of studies by \citet{PersicRephaeli2019FornaxA,PersicRephaeli2019II,PersicRephaeli2020III}.  
These works developed a uniform modeling approach for the broadband emission of radio lobes, demonstrating that the $\gamma$-ray output of several nearby systems can be explained by purely leptonic processes, without requiring a dominant hadronic component.  

The first analysis, focused on Fornax~A, reproduced its radio-to-$\gamma$-ray SED with a single truncated power-law electron distribution.  
The observed \textit{Fermi}-LAT flux was fully accounted for by IC scattering of the optical photons of NGC 1316 off the lobe electrons, implying magnetic fields of a few $\mu$G and confirming that the emission is predominantly leptonic \citep{PersicRephaeli2019FornaxA}.  
Subsequent work extended the same treatment to the lobes of Centaurus~A, Centaurus~B, and NGC~6251, showing that their $\gamma$-ray spectra are likewise produced by IC scattering of the combined CMB, EBL, and GFL \citep{PersicRephaeli2019II}.  

The third paper \citep{PersicRephaeli2020III} applied the model to six more distant FR\,I and FR\,II systems (3C\,98, Pictor~A, DA\,240, Cygnus~A, 3C\,326, 3C\,236) to predict their undetected IC components.  
Together, these studies show that the high-energy emission of radio lobes is mainly leptonic and that uniform SED modeling can robustly constrain magnetic fields, electron spectra, and upper limits on the proton energy density.

The unprecedented sensitivity and surface-brightness fidelity of SKA-Low and SKA-Mid will allow detailed mapping of the low-energy electron population and magnetic-field variations across radio lobes, anchoring future multi-wavelength modeling of their IC emission \citep[e.g.,][]{LalKraft2010,2011MNRAS.415..133H,LalKraft2013}.  
As demonstrated for the FR\,II galaxy IGR~J18249$-$3243 \citep{Bruni2022}, where the lobes dominate the $\gamma$-ray output (see Fig.~\ref{fig:IGR}), such joint radio and high-energy analyses will become routine with SKA and its VLBI extensions.

\subsection{Thermal and Coronal Emission}

In some FR\,I and FR\,II systems, particularly those with strong X-ray cores, thermal Comptonization in a hot accretion-disk corona contributes significantly in the keV range and may overlap with the non-thermal jet emission.  
The combination of coronal and SSC components explains the smooth connection between the hard X-ray and $\gamma$-ray bands observed in several misaligned AGN \citep{Paliya2024}.  
These coronal signatures also provide indirect evidence for a coupling between the accretion flow and the base of the jet, where energy dissipation transitions from thermal to non-thermal regimes.

\subsection{Hadronic and Hybrid Scenarios}

Although leptonic models (synchrotron, SSC, EC) account for most of the observed properties, hadronic interactions remain a plausible contributor in environments with dense photon fields or efficient particle acceleration.  
Proton–photon ($p\gamma$) interactions and proton-synchrotron emission can produce high-energy photons and neutrinos, albeit at a higher energetic cost.  
Given their large space density and moderate jet powers, FR\,0s have been proposed as potential contributors to the diffuse neutrino background \citep{Tavecchio2018,Paliya2021}. Even if each source contributes weakly, their collective output could become non-negligible on cosmic scales. 
In hybrid lepto-hadronic frameworks, the relative dominance of these channels depends on the baryon loading and magnetization of the jet plasma, parameters that future multi-messenger observations will help to constrain.

%\subsection{Synthesis}

%Across all misaligned radio-loud AGN classes, the $\gamma$-ray emission is thus shaped by a combination of synchrotron and Compton processes operating at different scales:
%\begin{itemize}
%    \item \textbf{Compact regions (parsec-scale):} SSC dominates, possibly aided by EC or structured-jet effects.
%    \item \textbf{Extended regions (kpc–Mpc):} IC on CMB/EBL photons becomes the primary high-energy mechanism.
%    \item \textbf{Thermal environments:} coronal Comptonization and weak EC from host photons can bridge to the X-ray domain.
%\end{itemize}
%This multi-scale, multi-process picture links young compact FR\,0s and CSOs with evolved FR\,IIs and GRGs, providing a coherent framework for the origin of high-energy emission in misaligned AGN.

%%%%%%%%%%%%%%%%%%%%%%%%%%%%%%%%%%%%%%%%%%%%%%%%%%%%%%%%%%%%%%%%

\section{Prospects with the SKA for high-energy radio galaxies}

The SKA will substantially tighten the physical picture emerging from \textit{Fermi}-era studies of misaligned AGN (FR\,0/FR\,I/FR\,II, GRGs, and CSOs) by providing uniform, high-fidelity radio constraints on jet energetics, environments, and particle distributions across two complementary regimes. 

\subsection{SKA1-Low}
SKA1-Low (50–350\,MHz) will probe optically thin, aging plasma and absorption turnovers with angular resolutions of $\sim$11$^{\prime\prime}$ at 110\,MHz and $\sim$4$^{\prime\prime}$ at 300\,MHz, and continuum sensitivities of $\sim$14–26\,$\mu$Jy\,beam$^{-1}$\,hr$^{-1}$ (for fractional bandwidth $\simeq0.3$) in the AA4 baseline configuration \citep{Dewdney2022}. At a typical redshift of the $\gamma$-ray-detected sample ($z\approx 0.05$), the SKA1-Low beam corresponds to physical scales of $\sim$5--10\,kpc. This will (i) map diffuse lobes and bridge emission in FR\,I/FR\,IIs and GRGs, enabling robust spectral-curvature/aging fits (break frequencies $\nu_{\rm br}$, radiative ages $t_{\rm rad}$) and energy closure (i.e., consistency between the observed radio synchrotron luminosity and the predicted inverse-Compton $\gamma$-ray output); (ii) separate low-frequency free–free absorption vs.\ synchrotron self-absorption in compact classes (FR\,0s, CSOs), localizing turnovers and constraining ambient densities on $\lesssim$\,kpc scales; and (iii) quantify low-energy electron populations that dominate external Compton predictions used in recent $\gamma$-ray modeling of young/compact sources. SKA1-Low’s surface-brightness sensitivity and wide fields will reveal Mpc-scale relic plasma and restarted episodes that are otherwise invisible, closing the energy budget needed to test lobe–IC $\gamma$-ray scenarios suggested for some misaligned AGN. 

\subsection{SKA1-Mid}
SKA1-Mid (0.35–15.4\,GHz) brings sub-arcsecond imaging (max resolutions $\sim$0.7$^{\prime\prime}$ at 0.77\,GHz, $\sim$0.4$^{\prime\prime}$ at 1.4\,GHz, $\sim$0.08$^{\prime\prime}$ at 6.7\,GHz, and $\sim$0.04$^{\prime\prime}$ at 12.5\,GHz) with continuum rms of $\sim$1.2–4.4\,$\mu$Jy beam$^{-1}$\,hr$^{-1}$ in the AA4 baseline configuration \citep{Dewdney2022}. At the same typical redshift ($z\approx 0.05$), SKA1-Mid reaches physical scales of $\sim$0.05--1\,kpc. This will (i) resolve cores, inner jets, and mini-lobes in FR\,0s and CSOs, enabling core–lobe decomposition, brightness-temperature limits, and precise SED anchoring for SSC vs.\ EC models used to interpret GeV detections and the occasional flaring behavior in young sources; (ii) deliver broad-band, matched-resolution spectra to measure in-situ acceleration (curved vs.\ broken power laws) and pinpoint total/partial opacity along jets; (iii) provide dense Faraday rotation grids and depolarization diagnostics to map magnetized circumnuclear gas and cluster atmospheres around FR\,I/FR\,II galaxies—key inputs for spine–sheath or multi-zone $\gamma$-ray models; and (iv) trace weak, restarted inner doubles inside large lobes, connecting duty cycle to high-energy variability. For GRGs, SKA1-Mid’s $\lesssim$0.1$^{\prime\prime}$ imaging at several GHz will isolate faint cores within bright, aged lobes, constraining contemporaneous jet power that feeds any observed high-energy component. 

In combination, \emph{Low}+{\em Mid} will provide spatially resolved, multi-frequency spectral curvature maps and polarization tomography from $\sim$100\,MHz to $>$10\,GHz, breaking the degeneracies between aging, absorption, and mixing that limit current models. This directly sharpens the inputs and priors adopted in recent population and SED studies of FR\,0s/CSOs/GRGs (e.g., compact-lobe IC efficiency, ambient photon fields, intermittent jet episodes) and enables targeted, contemporaneous tests with \textit{Fermi}/CTAO. Ultimately, SKA’s sensitivity and resolution turn today’s single-zone, angle-averaged descriptions into spatially resolved energy budgets, tying radio particle spectra and magnetic fields to the observed GeV output across all misaligned AGN subclasses.

\subsection{SKA-VLBI}

The inclusion of the SKA within global VLBI networks will revolutionize our understanding of the multi-scale structure and energetics of misaligned AGN, enabling a direct connection between radio morphologies, jet dynamics, and high-energy emission processes.  
In SKA-VLBI mode, the phased-array cores of SKA1-Mid and SKA1-Low will act as ultra-sensitive tied-array elements within existing VLBI networks (EVN, LBA, AVN), improving both the sensitivity and the $(u,v)$ coverage by orders of magnitude.

\subsubsection{Resolving the Jet–Emission Connection}

At frequencies from 1–15\,GHz (SKA1-Mid Bands 2–5), SKA-VLBI will achieve angular resolutions of $\sim$0.1–1\,mas, corresponding to linear scales of $\sim$0.1–1\,pc at the typical distances of $\gamma$-ray–detected radio galaxies ($z\lesssim0.1$).  
This is sufficient to spatially resolve the radio cores of FR\,0s, FR\,Is, and young CSOs, localizing the possible sites of particle acceleration and Compton up-scattering responsible for the observed GeV radiation.  
In particular, the combination of milliarcsecond imaging and broadband spectral coverage will allow:
i) direct mapping of brightness-temperature gradients and spectral turnovers along the inner jets, distinguishing between synchrotron and SSC-dominated zones;
ii) detection of compact sub-components (knots or mini-lobes) where magnetic reconnection or spine–sheath interactions could produce localized $\gamma$-ray flares;
iii) measurement of jet opening angles, Doppler factors, and apparent speeds from proper motion analysis for mildly misaligned jets, constraining the relativistic beaming corrections applied to high-energy luminosities.

For FR\,0 galaxies, which often remain unresolved in current VLBI surveys, SKA-VLBI will reveal whether their radio cores are truly compact or composed of multiple weak jetlets and shocks.  
This information is essential to test the SSC and EC models invoked to explain their $\gamma$-ray emission \citep{Paliya2021,Pannikkote2023}.

\subsubsection{Tracing Jet Evolution and Duty Cycle}

SKA-VLBI will bridge parsec- and kiloparsec-scale structures by combining its high-resolution core imaging with SKA-only interferometric maps at sub-arcsecond scales.  
This will provide a seamless view of jet collimation, disruption, and recurrence in FR\,Is and FR\,IIs (see also \citealt{hardcastle26SKA}).  
By comparing VLBI-scale cores with extended lobes detected by SKA-Mid and SKA-Low, it will be possible to determine: 
i) the energy transfer efficiency from the nuclear jet to the large-scale lobes by comparing the kinetic power inferred from VLBI jet kinematics ($P_{\rm jet,kin}$) with the total energy stored in the lobes (from SKA-Mid/Low integrated synchrotron luminosity and minimum-energy arguments) and with the lobe IC $\gamma$-ray luminosity (tested against NewAthena and CTAO data);
ii) the presence of restarted or intermittent jets in GRGs and FR\,0s, key for interpreting multi-epoch variability and spectral curvature in the high-energy band;
iii) the kinematic ages and expansion rates of young CSOs, directly connecting the early expansion phase to the onset of detectable $\gamma$-ray emission;
iv) the mechanisms of jet reorientation and precession by studying the relative orientation of jets on pc scales and of lobes/jets on kpc scales across the full population of radio galaxies (so far only investigated for a handful of systems, see \citealt{1988ApJ...328..114P,Ubertosi2024}): jet precession allows individual jet knots/blobs emitted by the cores of misaligned radio galaxies to be relatively more pointed towards the observer and generate gamma-ray variability (as in 3C\,120, \citealt{2015ApJ...808..162C}).

\subsubsection{Polarimetry, Magnetic Fields, and Jet Composition}

The exquisite sensitivity of SKA-VLBI polarimetry (rms noise $\lesssim$ few $\mu$Jy\,beam$^{-1}$ in full Stokes) will allow detailed studies of the magnetic-field structure in misaligned AGN.  
Rotation measure synthesis and Faraday depolarization analyses will reveal whether the $\gamma$-ray–bright regions coincide with the sites of enhanced magnetization or strong shear between jet layers.  
This is crucial to discriminate between leptonic and hadronic acceleration channels, since hadronic emission scenarios require high magnetic energy densities and strong magneto-hydrodynamic turbulence \citep{Tavecchio2018}.  
In CSOs, SKA-VLBI will trace the magnetized environment confining the mini-lobes, testing the IC  \citep{Stawarz2008} and bremsstrahlung \citep{Kino2009} models proposed for their GeV emission.

\section{Multi-wavelength and multi-messenger synergies with future facilities} 

The forthcoming \textit{NewAthena} X-ray observatory \citep{Cruise2025,Nandra2013} and the Cherenkov Telescope Array Observatory (CTAO; \citealt{Hofmann2023}) will complement the SKA by providing the missing high-energy anchor necessary to close the broadband energy balance of misaligned AGN. With its unprecedented effective area ($\sim$1.4\,m$^2$ at 1\,keV), high-resolution spectroscopy from the X-ray Integral Field Unit (X-IFU; $\Delta E\simeq2.5$\,eV), and wide-field imaging from the Wide Field Imager (WFI; 40$^{\prime}$ FoV, $<$5$^{\prime\prime}$ PSF), \textit{NewAthena} will enable spatially resolved diagnostics of the thermal and non-thermal plasma components in radio galaxies up to redshift $z\!\sim\!1$ \citep{Nandra2013,Barcons2017}. Combined SKA-\textit{NewAthena} datasets will provide a powerful, multi-scale view of jet–ISM/IGM coupling and particle acceleration in different stages of AGN evolution.

For FR\,0 and CSO galaxies, the synergy will be decisive in separating jet-related non-thermal X-rays from nuclear and circumnuclear thermal emission. The X-IFU spectra will constrain the absorbing columns ($N_{\rm H}$) and plasma temperatures in the dense environments that confine compact jets, while SKA-Mid will trace the associated synchrotron turnover and free–free absorption at sub-arcsecond scales. Such joint modeling will directly test the IC and bremsstrahlung mechanisms proposed for $\gamma$-ray–bright CSOs (e.g., PKS\,1718$-$649 and DA\,362). For FR\,I/FR\,II galaxies, simultaneous radio and X-ray imaging will resolve the co-spatial shocks and hot spots in inner jets and lobes, linking synchrotron and IC components, and measuring the local magnetic field strength without equipartition assumptions (as recently demonstrated in the powerful intermediate FR\,I/II galaxy Hercules A by \citealt{Ubertosi2025}). In GRG, \textit{NewAthena} will map thermal emission from the surrounding group/cluster medium and any residual inverse-Compton X-rays from aged lobes detected by SKA-Low, providing calorimetric estimates of total jet energy deposition.

On larger scales, \textit{NewAthena}'s sensitivity to diffuse, low-surface-brightness X-ray gas ($\sim10^{-16}$\,erg cm$^{-2}$\,s$^{-1}$) will trace cavities, shocks, and thermal winds inflated by AGN outflows that SKA will image in the radio continuum. Time-resolved observations of compact or recurrent jets will reveal whether high-energy flares—such as those recently observed in CSOs—are accompanied by X-ray coronal or jet-base variability, a crucial step toward understanding feedback and intermittency. Ultimately, the combined SKA-\textit{NewAthena} view will deliver the first self-consistent, multi-phase energy inventory of misaligned AGN, connecting particle acceleration, magnetic field evolution, and energy dissipation from parsec to megaparsec scales.

The synergy between SKA-VLBI, \textit{NewAthena}, and CTAO will enable contemporaneous imaging of both the low-energy and high-energy counterparts of misaligned AGN.  
While SKA-VLBI resolves the synchrotron jet base and measures the magnetic field configuration, \textit{NewAthena} will probe coronal and thermal IC components, and CTAO will extend the SED to the TeV regime. 
While full-sky synergies are not possible, the southern CTAO array at Paranal (Chile, latitude $\approx -24.7^{\circ}$) and SKA1-Mid in the Karoo region (South Africa, latitude $\approx -30.7^{\circ}$) share excellent common visibility. Coordinated campaigns can simultaneously observe up to $\sim$50--60\% of the celestial sphere with high elevation ($>30^{\circ}$) for both facilities, providing nearly complete access to the southern sky ($\delta\lesssim +30^{\circ}$).
Additionally, coordinated observations with the next-generation neutrino telescopes IceCube-Gen2 at the South Pole \citep{IceCube-Gen2:2021}, which provides excellent sensitivity to the northern celestial sky ($\delta \gtrsim -5^\circ$), and KM3NeT in the Mediterranean Sea \citep{KM3NeT:2016}, which offers superior sensitivity to the southern sky including the Galactic Centre region, will provide decisive tests for hadronic scenarios involving FR\,0s and other misaligned AGN. 

%%%%%%%%%%%%%%%%%%%%%%%%%%%%%%%%%%%%%%%%%%%%%%%%%%%
% BIBLIOGRAPHY
%%%%%%%%%%%%%%%%%%%%%%%%%%%%%%%%%%%%%%%%%%%%%%%%%%%
\bibliographystyle{abbrvnat-maxbibnames4}
\bibliography{HE-RG}

\begin{thebibliography}{71}
\providecommand{\natexlab}[1]{#1}
\providecommand{\url}[1]{\texttt{#1}}
\expandafter\ifx\csname urlstyle\endcsname\relax
  \providecommand{\doi}[1]{doi: #1}\else
  \providecommand{\doi}{doi: \begingroup \urlstyle{rm}\Url}\fi

\bibitem[{Abdo} et~al.(2010)]{Abdo2010}
A.~A. {Abdo} et~al.
\newblock \emph{The Astrophysical Journal}, 720:\penalty0 912--922, 2010.
\newblock \doi{10.1088/0004-637X/720/1/912}.

\bibitem[{Ajello} et~al.(2020){Ajello}, {Angioni}, {Axelsson}, {Ballet}, {Barbiellini}, {Bastieri}, {Becerra Gonzalez}, {Bellazzini}, {Bissaldi}, {Bloom}, {Bonino}, {Bottacini}, {Bruel}, {Buson}, {Cafardo}, {Cameron}, {Cavazzuti}, {Chen}, {Cheung}, {Ciprini}, {Costantin}, {Cutini}, {D'Ammando}, {de la Torre Luque}, {de Menezes}, {de Palma}, {Desai}, {Di Lalla}, {Di Venere}, {Dom{\'\i}nguez}, {Dirirsa}, {Ferrara}, {Finke}, {Franckowiak}, {Fukazawa}, {Funk}, {Fusco}, {Gargano}, {Garrappa}, {Gasparrini}, {Giglietto}, {Giordano}, {Giroletti}, {Green}, {Grenier}, {Guiriec}, {Harita}, {Hays}, {Horan}, {Itoh}, {J{\'o}hannesson}, {Kovac'evic'}, {Krauss}, {Kreter}, {Kuss}, {Larsson}, {Leto}, {Li}, {Liodakis}, {Longo}, {Loparco}, {Lott}, {Lovellette}, {Lubrano}, {Madejski}, {Maldera}, {Manfreda}, {Mart{\'\i}-Devesa}, {Massaro}, {Mazziotta}, {Mereu}, {Meyer}, {Migliori}, {Mirabal}, {Mizuno}, {Monzani}, {Morselli}, {Moskalenko}, {Negro}, {Nemmen}, {Nuss}, {Ojha}, {Ojha}, {Omodei}, {Orienti}, {Orlando}, {Ormes}, {Paliya},
  {Pei}, {Pe{\~n}a-Herazo}, {Persic}, {Pesce-Rollins}, {Petrov}, {Piron}, {Poon}, {Principe}, {Rain{\`o}}, {Rando}, {Rani}, {Razzano}, {Razzaque}, {Reimer}, {Reimer}, {Schinzel}, {Serini}, {Sgr{\`o}}, {Siskind}, {Spandre}, {Spinelli}, {Suson}, {Tachibana}, {Thompson}, {Torres}, {Torresi}, {Troja}, {Valverde}, {van Zyl}, and {Yassine}]{Ajello2020}
M.~{Ajello} et al.
\newblock \emph{\apj}, 892\penalty0 (2):\penalty0 105, Apr. 2020.
\newblock \doi{10.3847/1538-4357/ab791e}.

\bibitem[{Baldi}(2023)]{baldi23}
R.~D. {Baldi}.
\newblock \emph{The Astronomy and Astrophysics Review}, 31:\penalty0 3, 2023.
\newblock \doi{10.1007/s00159-023-00148-3}.

\bibitem[{Baldi} and {Capetti}(2009)]{Baldi2009}
R.~D. {Baldi} and A.~{Capetti}.
\newblock \emph{\aap}, 508\penalty0 (2):\penalty0 603--614, Dec. 2009.
\newblock \doi{10.1051/0004-6361/200913021}.

\bibitem[{Baldi} et~al.(2018){Baldi}, {Capetti}, and {Massaro}]{baldi18}
R.~D. {Baldi}, A.~{Capetti}, and F.~{Massaro}.
\newblock \emph{Astronomy \& Astrophysics}, 609:\penalty0 A1, 2018.
\newblock \doi{10.1051/0004-6361/201731333}.

\bibitem[{Baldi} et~al.(2019{\natexlab{a}}){Baldi}, {Capetti}, and {Giovannini}]{baldi19a}
R.~D. {Baldi}, A.~{Capetti}, and G.~{Giovannini}.
\newblock \emph{\mnras}, 482\penalty0 (2):\penalty0 2294--2304, Jan. 2019{\natexlab{a}}.
\newblock \doi{10.1093/mnras/sty2703}.

\bibitem[{Baldi} et~al.(2019{\natexlab{b}}){Baldi}, {Capetti}, and {Giovannini}]{baldi19b}
R.~D. {Baldi}, A.~{Capetti}, and G.~{Giovannini}.
\newblock \emph{Monthly Notices of the Royal Astronomical Society}, 482:\penalty0 2294--2305, 2019{\natexlab{b}}.
\newblock \doi{10.1093/mnras/sty2848}.

\bibitem[{Baldi} et~al.(2019{\natexlab{c}}){Baldi}, {Torresi}, {Migliori}, {Balmaverde}, et~al.]{baldi19}
R.~D. {Baldi} et al.
\newblock \emph{Galaxies}, 7\penalty0 (3):\penalty0 76, 2019{\natexlab{c}}.
\newblock \doi{10.3390/galaxies7030076}.

\bibitem[Baldi et~al.(2026)Baldi, author2, author3, author4, and author5]{baldi26SKA}
R.~D. Baldi et al.
\newblock In \emph{Advancing Astrophysics with the SKA -- II (AASKAII)}. 2026.
\newblock arXiv search: Report number AASKAII/Baldi01.

\bibitem[{Barcons} et~al.(2017)]{Barcons2017}
X.~{Barcons} et~al.
\newblock \emph{Astronomy \& Astrophysics}, 608:\penalty0 A1, 2017.

\bibitem[{Becker} et~al.(1995){Becker}, {White}, and {Helfand}]{Becker1995}
R.~H. {Becker}, R.~L. {White}, and D.~J. {Helfand}.
\newblock \emph{\apj}, 450:\penalty0 559, Sept. 1995.
\newblock \doi{10.1086/176166}.

\bibitem[{Best} and {Heckman}(2012)]{Best2012}
P.~N. {Best} and T.~M. {Heckman}.
\newblock \emph{Monthly Notices of the Royal Astronomical Society}, 421:\penalty0 1569--1582, 2012.
\newblock \doi{10.1111/j.1365-2966.2012.20414.x}.

\bibitem[{Boughelilba} and {Reimer}(2023)]{2023ApJ...955L..41B}
M.~{Boughelilba} and A.~{Reimer}.
\newblock \emph{\apjl}, 955\penalty0 (2):\penalty0 L41, Oct. 2023.
\newblock \doi{10.3847/2041-8213/acf83c}.

\bibitem[{Bronzini} et~al.(2024){Bronzini}, {Grandi}, {Torresi}, and {Buson}]{Bronzini}
E.~{Bronzini}, P.~{Grandi}, E.~{Torresi}, and S.~{Buson}.
\newblock \emph{\apjl}, 977\penalty0 (1):\penalty0 L16, Dec. 2024.
\newblock \doi{10.3847/2041-8213/ad93cf}.

\bibitem[{Bruni} et~al.(2020){Bruni}, {Panessa}, {Bassani}, et~al.]{Bruni2020}
G.~{Bruni}, F.~{Panessa}, L.~{Bassani}, et~al.
\newblock \emph{Monthly Notices of the Royal Astronomical Society}, 494:\penalty0 902--914, 2020.
\newblock \doi{10.1093/mnras/staa735}.

\bibitem[{Bruni} et~al.(2022){Bruni}, {Panessa}, {Bassani}, et~al.]{Bruni2022}
G.~{Bruni}, F.~{Panessa}, L.~{Bassani}, et~al.
\newblock \emph{Monthly Notices of the Royal Astronomical Society}, 513:\penalty0 886--899, 2022.
\newblock \doi{10.1093/mnras/stac865}.

\bibitem[{Cao} et~al.(2024){Cao}, {Aharonian}, {Axikegu}, {Bai}, {Bao}, {Bastieri}, {Bi}, {Bi}, {Bian}, {Bukevich}, {Cao}, {Cao}, {Cao}, {Chang}, {Chang}, {Chen}, {Chen}, {Chen}, {Chen}, {Chen}, {Chen}, {Chen}, {Chen}, {Chen}, {Chen}, {Chen}, {Chen}, {Chen}, {Chen}, {Cheng}, {Cheng}, {Cui}, {Cui}, {Cui}, {Cui}, {Dai}, {Dai}, {Dai}, {Danzengluobu}, {Dong}, {Duan}, {Fan}, {Fan}, {Fang}, {Fang}, {Fang}, {Feng}, {Feng}, {Feng}, {Feng}, {Feng}, {Feng}, {Feng}, {Gabici}, {Gao}, {Gao}, {Gao}, {Gao}, {Gao}, {Ge}, {Geng}, {Giacinti}, {Gong}, {Gou}, {Gu}, {Guo}, {Guo}, {Guo}, {Guo}, {Han}, {Hasan}, {He}, {He}, {He}, {He}, {Hor}, {Hou}, {Hou}, {Hou}, {Hu}, {Hu}, {Hu}, {Huang}, {Huang}, {Huang}, {Huang}, {Huang}, {Huang}, {Ji}, {Jia}, {Jia}, {Jiang}, {Jiang}, {Jiang}, {Jin}, {Kang}, {Karpikov}, {Kuleshov}, {Kurinov}, {Li}, {Li}, {Li}, {Li}, {Li}, {Li}, {Li}, {Li}, {Li}, {Li}, {Li}, {Li}, {Li}, {Li}, {Li}, {Li}, {Li}, {Li}, {Li}, {Liang}, {Liang}, {Lin}, {Liu}, {Liu}, {Liu}, {Liu}, {Liu}, {Liu}, {Liu}, {Liu}, {Liu},
  {Liu}, {Liu}, {Liu}, {Liu}, {Liu}, {Luo}, {Luo}, {Lv}, {Ma}, {Ma}, {Ma}, {Mao}, {Min}, {Mitthumsiri}, {Mu}, {Nan}, {Neronov}, {Ou}, {Pattarakijwanich}, {Pei}, {Qi}, {Qi}, {Qiao}, {Qin}, {Raza}, {Ruffolo}, {S{\'a}iz}, {Saeed}, {Semikoz}, {Shao}, {Shchegolev}, {Sheng}, {Shu}, {Song}, {Stenkin}, {Stepanov}, {Su}, {Sun}, {Sun}, {Sun}, {Sun}, {Takata}, {Tam}, {Tang}, {Tang}, {Tang}, {Tian}, {Wang}, {Wang}, {Wang}, {Wang}, {Wang}, {Wang}, {Wang}, {Wang}, {Wang}, {Wang}, {Wang}, {Wang}, {Wang}, {Wang}, {Wang}, {Wang}, {Wang}, {Wang}, {Wang}, {Wang}, {Wang}, {Wang}, and {Wei}]{Cao}
Z.~{Cao} et al.
\newblock \emph{\apjl}, 971\penalty0 (2):\penalty0 L45, Aug. 2024.
\newblock \doi{10.3847/2041-8213/ad5e6d}.

\bibitem[{Casadio} et~al.(2015){Casadio}, {G{\'o}mez}, {Grandi}, {Jorstad}, {Marscher}, {Lister}, {Kovalev}, {Savolainen}, and {Pushkarev}]{2015ApJ...808..162C}
C.~{Casadio} et al.
\newblock \emph{\apj}, 808\penalty0 (2):\penalty0 162, Aug. 2015.
\newblock \doi{10.1088/0004-637X/808/2/162}.

\bibitem[{Condon} et~al.(1998){Condon}, {Cotton}, {Greisen}, {Yin}, {Perley}, {Taylor}, and {Broderick}]{Condon1998}
J.~J. {Condon} et al.
\newblock \emph{\aj}, 115\penalty0 (5):\penalty0 1693--1716, May 1998.
\newblock \doi{10.1086/300337}.

\bibitem[{Cruise} et~al.(2025){Cruise}, {Guainazzi}, {Aird}, et~al.]{Cruise2025}
M.~{Cruise}, M.~{Guainazzi}, J.~{Aird}, et~al.
\newblock \emph{Nature Astronomy}, 9:\penalty0 36--44, 2025.
\newblock \doi{10.1038/s41550-024-02416-3}.
\newblock arXiv:2501.03100.

\bibitem[{Dabhade} et~al.(2020){Dabhade}, {Mahato}, {Bagchi}, et~al.]{Dabhade2020}
P.~{Dabhade}, M.~{Mahato}, J.~{Bagchi}, et~al.
\newblock \emph{Astronomy \& Astrophysics}, 642:\penalty0 A153, 2020.
\newblock \doi{10.1051/0004-6361/202037049}.

\bibitem[Dewdney et~al.(2022)Dewdney, Labate, Swart, Stringhetti, McMullin, Alachkar, Martin, Bartolini, Bourke, Bolton, Breen, Bridges, Caiazzo, Chen, Chrysostomou, Cremonini, Di~Vruno, García~Miró, Faye~Hammond, Graser, Hayden, Hermann, Hendre, Jameson, Karastergiou, Lewis, Le~Roux, Lloyd, Nkwau, Obiebi, Olguin, Otto, Pellegrini, Roy, Schutte, Smith, Strydom, Swart, Taljaard, van Zyl, Wagg, Waterson, and Morgan]{Dewdney2022}
P.~Dewdney et al.
\newblock Ska1 design baseline description, Jan. 2022.
\newblock URL \url{https://doi.org/10.5281/zenodo.16895574}.

\bibitem[{Fanaroff} et~al.(2021){Fanaroff}, {Lal}, {Venturi}, {Smirnov}, {Bondi}, {Thorat}, {Bester}, {J{\'o}zsa}, {Kleiner}, {Loi}, {Makhathini}, and {White}]{Bernie2021}
B.~{Fanaroff} et al.
\newblock \emph{\mnras}, 505:\penalty0 6003--6016, 2021.
\newblock \doi{10.1093/mnras/stab1540}.

\bibitem[{Fanaroff} and {Riley}(1974)]{Fanaroff1974}
B.~L. {Fanaroff} and J.~M. {Riley}.
\newblock \emph{Monthly Notices of the Royal Astronomical Society}, 167:\penalty0 31P--36P, 1974.
\newblock \doi{10.1093/mnras/167.1.31P}.

\bibitem[{Ghisellini}(2011)]{Ghisellini2011}
G.~{Ghisellini}.
\newblock \emph{Memorie della Societ{\`a} Astronomica Italiana}, 82:\penalty0 104--113, 2011.

\bibitem[Ghisellini et~al.(2005)Ghisellini, Tavecchio, and Chiaberge]{ghisellini2005}
G.~Ghisellini, F.~Tavecchio, and M.~Chiaberge.
\newblock \emph{A\&A}, 432:\penalty0 401--410, 2005.
\newblock \doi{10.1051/0004-6361:20041404}.

\bibitem[{Grandi} et~al.(2016){Grandi}, {Capetti}, and {Baldi}]{Grandi2016}
P.~{Grandi}, A.~{Capetti}, and R.~D. {Baldi}.
\newblock \emph{Monthly Notices of the Royal Astronomical Society}, 457:\penalty0 2--12, 2016.
\newblock \doi{10.1093/mnras/stv2894}.

\bibitem[{Hardcastle} and {Croston}(2011)]{2011MNRAS.415..133H}
M.~J. {Hardcastle} and J.~H. {Croston}.
\newblock \emph{\mnras}, 415\penalty0 (1):\penalty0 133--142, July 2011.
\newblock \doi{10.1111/j.1365-2966.2011.18678.x}.

\bibitem[{Hardcastle} et~al.(2019){Hardcastle}, {Williams}, {Best}, {Croston}, {Duncan}, {R{\"o}ttgering}, {Sabater}, {Shimwell}, {Tasse}, {Callingham}, {Cochrane}, {de Gasperin}, {G{\"u}rkan}, {Jarvis}, {Mahatma}, {Miley}, {Mingo}, {Mooney}, {Morabito}, {O'Sullivan}, {Prandoni}, {Shulevski}, and {Smith}]{Hardcastle2019}
M.~J. {Hardcastle} et al.
\newblock \emph{\aap}, 622:\penalty0 A12, Feb. 2019.
\newblock \doi{10.1051/0004-6361/201833893}.

\bibitem[Hardcastle et~al.(2026)Hardcastle, author2, author3, author4, and author5]{hardcastle26SKA}
M.~J. Hardcastle et al.
\newblock In \emph{Advancing Astrophysics with the SKA -- II (AASKAII)}. 2026.
\newblock arXiv search: Report number AASKAII/Hardcastle01.

\bibitem[{Hofmann} et~al.(2023){Hofmann}, {Zanin}, and {CTA Consortium}]{Hofmann2023}
W.~{Hofmann}, R.~{Zanin}, and {CTA Consortium}.
\newblock \emph{arXiv e-prints}, art. arXiv:2305.12888, 2023.

\bibitem[{IceCube-Gen2 Collaboration}(2021)]{IceCube-Gen2:2021}
{IceCube-Gen2 Collaboration}.
\newblock \emph{J. Phys. G}, 48\penalty0 (6):\penalty0 060501, 2021.
\newblock \doi{10.1088/1361-6471/abbd48}.

\bibitem[{Khatiya} et~al.(2024){Khatiya}, {Boughelilba}, {Karwin}, {McDaniel}, {Zhao}, {Ajello}, {Reimer}, and {Hartmann}]{2024ApJ...971...84K}
N.~S. {Khatiya} et al.
\newblock \emph{\apj}, 971\penalty0 (1):\penalty0 84, Aug. 2024.
\newblock \doi{10.3847/1538-4357/ad534c}.

\bibitem[{Kino} and {Asano}(2011)]{KinoAsano2011}
M.~{Kino} and K.~{Asano}.
\newblock \emph{Monthly Notices of the Royal Astronomical Society: Letters}, 412:\penalty0 L20--L24, 2011.
\newblock \doi{10.1111/j.1745-3933.2010.00999.x}.

\bibitem[{Kino} et~al.(2009){Kino}, {Ito}, {Kawakatu}, and {Nagai}]{Kino2009}
M.~{Kino}, H.~{Ito}, N.~{Kawakatu}, and H.~{Nagai}.
\newblock \emph{\mnras}, 395\penalty0 (1):\penalty0 L43--L47, May 2009.
\newblock \doi{10.1111/j.1745-3933.2009.00638.x}.

\bibitem[{KM3NeT Collaboration}(2016)]{KM3NeT:2016}
{KM3NeT Collaboration}.
\newblock \emph{J. Phys. G}, 43\penalty0 (8):\penalty0 084001, 2016.
\newblock \doi{10.1088/0954-3899/43/8/084001}.

\bibitem[{Lal}(2021)]{Lalremnant2021}
D.~V. {Lal}.
\newblock \emph{\apj}, 915\penalty0 (2):\penalty0 126, 2021.
\newblock \doi{10.3847/1538-4357/ac042d}.

\bibitem[{Lal} and {Rao}(2004)]{LalRao04}
D.~V. {Lal} and A.~P. {Rao}.
\newblock \emph{\aap}, 420:\penalty0 491--499, 2004.
\newblock \doi{10.1051/0004-6361:20035777}.

\bibitem[{Lal} et~al.(2008){Lal}, {Hardcastle}, and {Kraft}]{LalHardcastle2008}
D.~V. {Lal}, M.~J. {Hardcastle}, and R.~P. {Kraft}.
\newblock \emph{\mnras}, 390\penalty0 (3):\penalty0 1105--1116, 2008.
\newblock \doi{10.1111/j.1365-2966.2008.13810.x}.

\bibitem[{Lal} et~al.(2010){Lal}, {Kraft}, {Forman}, {Hardcastle}, {Jones}, {Nulsen}, {Evans}, {Croston}, and {Lee}]{LalKraft2010}
D.~V. {Lal} et al.
\newblock \emph{\apj}, 722:\penalty0 1735--1743, Oct. 2010.
\newblock \doi{10.1088/0004-637X/722/2/1735}.

\bibitem[{Lal} et~al.(2013){Lal}, {Kraft}, {Randall}, {Forman}, {Nulsen}, {Roediger}, {ZuHone}, {Hardcastle}, {Jones}, and {Croston}]{LalKraft2013}
D.~V. {Lal} et al.
\newblock \emph{\apj}, 764:\penalty0 83, 2013.
\newblock \doi{10.1088/0004-637X/764/1/83}.

\bibitem[{Lal} et~al.(2021){Lal}, {Legodi}, {Fanaroff}, {Venturi}, {Smirnov}, {Bondi}, {Thorat}, {Bester}, {J{\'o}zsa}, {Kleiner}, {Loi}, {Makhathini}, and {White}]{LalBernie2021}
D.~V. {Lal} et al.
\newblock \emph{Galaxies}, 9:\penalty0 87, 2021.
\newblock \doi{10.3390/galaxies9040087}.

\bibitem[{Lister} et~al.(2020){Lister}, {Homan}, {Kovalev}, {Kellermann}, {Richards}, {Ros}, {Meyer}, {Kovalev}, {Savolainen}, et~al.]{Lister2020}
M.~L. {Lister} et al.
\newblock \emph{The Astrophysical Journal}, 899:\penalty0 141, 2020.
\newblock \doi{10.3847/1538-4357/aba18d}.

\bibitem[{McConnell} et~al.(2020){McConnell}, {Hale}, {Lenc}, {Banfield}, {Heald}, {Hotan}, {Leung}, {Moss}, {Murphy}, {O'Brien}, {Pritchard}, {Raja}, {Sadler}, {Stewart}, {Thomson}, {Whiting}, {Allison}, {Amy}, {Anderson}, {Ball}, {Bannister}, {Bell}, {Bock}, {Bolton}, {Bunton}, {Chippendale}, {Collier}, {Cooray}, {Cornwell}, {Diamond}, {Edwards}, {Gupta}, {Hayman}, {Heywood}, {Jackson}, {Koribalski}, {Lee-Waddell}, {McClure-Griffiths}, {Ng}, {Norris}, {Phillips}, {Reynolds}, {Roxby}, {Schinckel}, {Shields}, {Tremblay}, {Tzioumis}, {Voronkov}, and {Westmeier}]{McConnell2020}
D.~{McConnell} et al.
\newblock \emph{\pasa}, 37:\penalty0 e048, Nov. 2020.
\newblock \doi{10.1017/pasa.2020.41}.

\bibitem[{Migliori} et~al.(2014){Migliori}, {Siemiginowska}, {Kelly}, {Stawarz}, {Celotti}, and {Begelman}]{Migliori2014}
G.~{Migliori} et al.
\newblock \emph{The Astrophysical Journal}, 780:\penalty0 165, 2014.
\newblock \doi{10.1088/0004-637X/780/2/165}.

\bibitem[{Migliori} et~al.(2016){Migliori}, {Siemiginowska}, {Sobolewska}, {Loh}, {Corbel}, {Ostorero}, and {Stawarz}]{Migliori2016}
G.~{Migliori} et al.
\newblock \emph{The Astrophysical Journal Letters}, 821:\penalty0 L31, 2016.
\newblock \doi{10.3847/2041-8205/821/2/L31}.

\bibitem[{Mingo} et~al.(2019){Mingo}, {Croston}, {Hardcastle}, {Best}, {Duncan}, {Morganti}, {Rottgering}, {Sabater}, {Shimwell}, {Williams}, {Brienza}, {Gurkan}, {Mahatma}, {Morabito}, {Prandoni}, {Bondi}, {Ineson}, and {Mooney}]{Mingo2019}
B.~{Mingo} et al.
\newblock \emph{\mnras}, 488\penalty0 (2):\penalty0 2701--2721, Sept. 2019.
\newblock \doi{10.1093/mnras/stz1901}.

\bibitem[{Mingo} et~al.(2022){Mingo}, {Croston}, {Best}, {Duncan}, {Hardcastle}, {Kondapally}, {Prandoni}, {Sabater}, {Shimwell}, {Williams}, {Baldi}, {Bonato}, {Bondi}, {Dabhade}, {G{\"u}rkan}, {Ineson}, {Magliocchetti}, {Miley}, {Pierce}, and {R{\"o}ttgering}]{Mingo2022}
B.~{Mingo} et al.
\newblock \emph{\mnras}, 511\penalty0 (3):\penalty0 3250--3271, Apr. 2022.
\newblock \doi{10.1093/mnras/stac140}.

\bibitem[{Nandra} et~al.(2013){Nandra}, {Barret}, {Barcons}, {Fabian}, {den Herder}, {Piro}, {Watson}, et~al.]{Nandra2013}
K.~{Nandra} et al.
\newblock Athena mission proposal, 2013.

\bibitem[{O'Dea}(1998)]{ODea1998}
C.~P. {O'Dea}.
\newblock \emph{Publications of the Astronomical Society of the Pacific}, 110:\penalty0 493--532, 1998.
\newblock \doi{10.1086/316162}.

\bibitem[{Ostorero} et~al.(2010){Ostorero}, {Moderski}, {Stawarz}, {Diaferio}, {Kowalska}, {Cheung}, {Kataoka}, {Begelman}, and {Wagner}]{Ostorero2010}
L.~{Ostorero} et al.
\newblock \emph{\apj}, 715\penalty0 (2):\penalty0 1071--1093, June 2010.
\newblock \doi{10.1088/0004-637X/715/2/1071}.

\bibitem[{Paliya}(2021)]{Paliya2021}
V.~S. {Paliya}.
\newblock \emph{The Astrophysical Journal Letters}, 918:\penalty0 L39, 2021.
\newblock \doi{10.3847/2041-8213/ac1f0d}.

\bibitem[{Paliya} et~al.(2024){Paliya}, {Saikia}, {Dom{\'\i}nguez}, {Stalin}, et~al.]{Paliya2024}
V.~S. {Paliya} et al.
\newblock \emph{The Astrophysical Journal}, 976:\penalty0 120, 2024.
\newblock \doi{10.3847/1538-4357/ad85e2}.

\bibitem[{Paliya} et~al.(2025)]{Paliya2025}
V.~S. {Paliya} et~al.
\newblock \emph{The Astrophysical Journal}, 989:\penalty0 36, 2025.

\bibitem[{Pannikkote} et~al.(2023){Pannikkote}, {Paliya}, and {Saikia}]{Pannikkote2023}
M.~{Pannikkote}, V.~S. {Paliya}, and D.~J. {Saikia}.
\newblock \emph{The Astrophysical Journal}, 957:\penalty0 73, 2023.
\newblock \doi{10.3847/1538-4357/acf6aa}.

\bibitem[{Pearson} and {Readhead}(1988)]{1988ApJ...328..114P}
T.~J. {Pearson} and A.~C.~S. {Readhead}.
\newblock \emph{\apj}, 328:\penalty0 114, May 1988.
\newblock \doi{10.1086/166274}.

\bibitem[{Persic} and {Rephaeli}(2019{\natexlab{a}})]{PersicRephaeli2019FornaxA}
M.~{Persic} and Y.~{Rephaeli}.
\newblock \emph{Monthly Notices of the Royal Astronomical Society}, 485:\penalty0 2001--2009, 2019{\natexlab{a}}.
\newblock \doi{10.1093/mnras/stz511}.

\bibitem[{Persic} and {Rephaeli}(2019{\natexlab{b}})]{PersicRephaeli2019II}
M.~{Persic} and Y.~{Rephaeli}.
\newblock \emph{Monthly Notices of the Royal Astronomical Society}, 490:\penalty0 1489--1502, 2019{\natexlab{b}}.
\newblock \doi{10.1093/mnras/stz2680}.

\bibitem[{Persic} and {Rephaeli}(2020)]{PersicRephaeli2020III}
M.~{Persic} and Y.~{Rephaeli}.
\newblock \emph{Monthly Notices of the Royal Astronomical Society}, 491:\penalty0 5740--5757, 2020.
\newblock \doi{10.1093/mnras/stz3659}.

\bibitem[{Principe} et~al.(2020){Principe}, {Migliori}, {Johnson}, {Lico}, {Orienti}, et~al.]{Principe2020}
G.~{Principe} et al.
\newblock \emph{Astronomy \& Astrophysics}, 635:\penalty0 A185, 2020.
\newblock \doi{10.1051/0004-6361/201937049}.

\bibitem[{Readhead} et~al.(1996){Readhead}, {Taylor}, {Xu}, {Pearson}, {Wilkinson}, and {Polatidis}]{Readhead1996}
A.~C.~S. {Readhead} et al.
\newblock \emph{The Astrophysical Journal}, 460:\penalty0 612--627, 1996.
\newblock \doi{10.1086/176996}.

\bibitem[{Shimwell} et~al.(2017){Shimwell}, {R{\"o}ttgering}, {Best}, {Williams}, {Dijkema}, {de Gasperin}, {Hardcastle}, {Heald}, {Hoang}, {Horneffer}, {Intema}, {Mahony}, {Mandal}, {Mechev}, {Morabito}, {Oonk}, {Rafferty}, {Retana-Montenegro}, {Sabater}, {Tasse}, {van Weeren}, {Br{\"u}ggen}, {Brunetti}, {Chy{\.z}y}, {Conway}, {Haverkorn}, {Jackson}, {Jarvis}, {McKean}, {Miley}, {Morganti}, {White}, {Wise}, {van Bemmel}, {Beck}, {Brienza}, {Bonafede}, {Calistro Rivera}, {Cassano}, {Clarke}, {Cseh}, {Deller}, {Drabent}, {van Driel}, {Engels}, {Falcke}, {Ferrari}, {Fr{\"o}hlich}, {Garrett}, {Harwood}, {Heesen}, {Hoeft}, {Horellou}, {Israel}, {Kapi{\'n}ska}, {Kunert-Bajraszewska}, {McKay}, {Mohan}, {Orr{\'u}}, {Pizzo}, {Prandoni}, {Schwarz}, {Shulevski}, {Sipior}, {Smith}, {Sridhar}, {Steinmetz}, {Stroe}, {Varenius}, {van der Werf}, {Zensus}, and {Zwart}]{Shimwell2017}
T.~W. {Shimwell} et al.
\newblock \emph{\aap}, 598:\penalty0 A104, Feb. 2017.
\newblock \doi{10.1051/0004-6361/201629313}.

\bibitem[{Stawarz} et~al.(2008){Stawarz}, {Ostorero}, {Begelman}, {Moderski}, {Kataoka}, and {Wagner}]{Stawarz2008}
{\L}.~{Stawarz} et al.
\newblock \emph{\apj}, 680\penalty0 (2):\penalty0 911--925, June 2008.
\newblock \doi{10.1086/587781}.

\bibitem[{Swain} et~al.(2025){Swain}, {Paliya}, {Saikia}, and {Stalin}]{Swain2025}
S.~{Swain}, V.~S. {Paliya}, D.~J. {Saikia}, and C.~S. {Stalin}.
\newblock \emph{The Astrophysical Journal}, 979:\penalty0 97, 2025.

\bibitem[Tavecchio and Ghisellini(2008)]{tavecchio2008}
F.~Tavecchio and G.~Ghisellini.
\newblock \emph{MNRAS Letters}, 385:\penalty0 L98--L102, 2008.
\newblock \doi{10.1111/j.1745-3933.2008.00441.x}.

\bibitem[{Tavecchio} et~al.(2018){Tavecchio}, {Righi}, {Capetti}, {Grandi}, and {Ghisellini}]{Tavecchio2018}
F.~{Tavecchio} et al.
\newblock \emph{Monthly Notices of the Royal Astronomical Society}, 475:\penalty0 5529--5534, 2018.
\newblock \doi{10.1093/mnras/sty251}.

\bibitem[{Torresi} et~al.(2018){Torresi}, {Grandi}, {Capetti}, {Baldi}, and {Giovannini}]{torresi18}
E.~{Torresi} et al.
\newblock \emph{Monthly Notices of the Royal Astronomical Society}, 476:\penalty0 5535--5547, 2018.
\newblock \doi{10.1093/mnras/sty520}.

\bibitem[{Ubertosi} et~al.(2024){Ubertosi}, {Schellenberger}, {O'Sullivan}, {Vrtilek}, {Giacintucci}, {David}, {Forman}, {Gitti}, {Venturi}, {Jones}, and {Brighenti}]{Ubertosi2024}
F.~{Ubertosi} et al.
\newblock \emph{\apj}, 961\penalty0 (1):\penalty0 134, Jan. 2024.
\newblock \doi{10.3847/1538-4357/ad11d8}.

\bibitem[{Ubertosi} et~al.(2025){Ubertosi}, {Gong}, {Nulsen}, {Leahy}, {Gitti}, {McNamara}, {Gaspari}, {Singha}, {O'Dea}, and {Baum}]{Ubertosi2025}
F.~{Ubertosi} et al.
\newblock \emph{\aap}, 693:\penalty0 A171, Jan. 2025.
\newblock \doi{10.1051/0004-6361/202452430}.

\bibitem[{Wilkinson} et~al.(1994){Wilkinson}, {Polatidis}, {Readhead}, {Xu}, and {Pearson}]{Wilkinson1994}
P.~N. {Wilkinson} et al.
\newblock \emph{The Astrophysical Journal Letters}, 432:\penalty0 L87--L90, 1994.
\newblock \doi{10.1086/187516}.

\bibitem[{Ye} et~al.(2026){Ye}, {Baldi}, {Yang}, {Zhu}, {Bastieri}, {Bachev}, {Strigachev}, and {Fan}]{ye26}
X.-H. {Ye} et al.
\newblock \emph{\aap}, 708:\penalty0 A56, Mar. 2026.
\newblock \doi{10.1051/0004-6361/202556939}.

\end{thebibliography}

%%%%%%%%%%%%%%%%%%%%%%%%%%%%%%%%%%%%%%%%%%%%%%%%%%%
%%%%%%%%%%%%%%%%%%%%%%%%%%%%%%%%%%%%%%%%%%%%%%%%%%%
\end{document}